\documentclass[10.5pt]{article}

\usepackage{scicite}
\usepackage{float}
\usepackage{graphicx}
\usepackage{times}

\newenvironment{sciabstract}{%
\begin{quote} \bf}
{\end{quote}}

\title{Quantum thermodynamics with a single superconducting vortex} 

\author
{Marek Foltyn,$^{1}$ Konrad Norowski,$^{1}$ Alexander Savin,$^{2}$ Maciej Zgirski$^{1\ast}$\\
\\
\normalsize{$^{1}$Institute of Physics, Polish Academy of Sciences,}\\
\normalsize{Aleja Lotnikow 32/46, Warsaw, PL 02668, Poland}\\
\normalsize{$^{2}$QTF Centre of Excellence, Department of Applied Physics,}\\
\normalsize{Aalto University, Aalto, FI-00076, Finland}\\
\\
\normalsize{$^\ast$To whom correspondence should be addressed; E-mail: zgirski@ifpan.edu.pl}
}

\date{}

\begin{document} 

\baselineskip24pt

\maketitle

\begin{sciabstract}
We demonstrate complete control over dynamics of a single superconducting vortex in a nanostructure which we coin the Single Vortex Box (SVB). Our device allows us to trap the vortex in a field-cooled aluminum nanosquare and expel it on demand with a nanosecond pulse of electrical current. We read-out the vortex state of the box by testing the switching current of the adjacent Dayem nanobridge. Using the time–resolving nanothermometry we measure 4$\cdot$10$^{-19}\,$J as the amount of the dissipated heat (which is the energy of a single red photon) in the elementary process of the vortex expulsion, and monitor the following thermal relaxation of the device. The measured heat is equal to the energy required to annihilate all Cooper pairs on the way of the moving vortex. Our design and measuring protocol are convenient for studying the stochastic mechanism of the vortex escape from  current-driven superconducting nanowires, which has its roots either in thermal or quantum fluctuations, similar to ones widely studied in Josephson junctions or magnetic nanoclusters and molecules. Our experiment enlightens the thermodynamics of the absorption process in the superconducting nanowire single-photon detectors, in which vortices are perceived to be essential for a formation of a detectable hot spot. The demonstrated opportunity to manipulate a single superconducting vortex reliably in a confined geometry comprises in fact a proof-of-concept of a nanoscale non-volatile memory cell with sub-nanosecond write and read operations, which offers compatibility with quantum processors based either on superconducting qubits or rapid single flux quantum circuits.
\end{sciabstract}

\section*{Introduction}

Thermodynamics involves studies of the heat flow arising from the difference in temperature between two bodies, as stated in the 2$^{nd}$ law. When such flow is considered at a single undividable particle level, we investigate the thermodynamics at its natural microscopic limit governed by quantum physics\cite{Pekola2015}. Such studies are preferably performed in nanoscale devices cooled down to the lowest temperature where the quantum effects can flourish and temperature gradients can be set on demand. The seminal experiments performed in the field involve demonstrations of quantized thermal conductance of heat not only by single modes of phonons\cite{Schwab2000} and photons\cite{Meschke2006,Partanen2016} but also single electron channels\cite{Jezouin2013} and anyons\cite{Banerjee2017}. Besides the steady-state investigations researchers were able to demonstrate the control of the heat transport at a single particle level as exemplified in the experiment with an electron turnstile\cite{Marin2022}. Interestingly, the laws of thermodynamics first written down in 19th century, owing to their only statistical validity, do not need to hold for microscopic systems which exchange quantized amount of energy, e.g. it is possible to observe heat flow from colder to hotter object albeit with lower probability than in the opposite direction favoured by the 2$^{nd}$ law of thermodynamics\cite{Utsumi2010,Pekola2015}. The monitoring and control of the heat transport at a single particle level allowed researchers recently to revive the Maxwell demon, who for long time seemed to be only an intellectual curiosity\cite{Koski2015,Maruyama2009}. The experimental verification of the Landauer's principle linking the erasure of a single bit of information with the minimum amount of dissipated heat of $k_B T ln(2)$ (i.e. the Landauer bound) exorcised the demon and connected two worlds: the information theory and thermodynamics\cite{Lutz2012}.

The recent advancements in experimental techniques, particularly in nanothermometry, have allowed to shine a new light on various frequently studied quantum phenomena, in which role of dissipation had been only postulated, sometimes a priori neglected, but never verified experimentally. Researchers were able to perform thermal imaging of a graphene with SQUID-on-tip and found the dissipation in resonant states along the edges of the sample\cite{Halbertal2016}. The other team measured a pronounced temperature rise in a nanoscopic metallic island serving as a junction in an RF-SQUIPT due to a single phase slip event\cite{Winkelmann2023}. Similar experiments are expected to deeply affect our understanding of the dynamics of the quantum systems, in which dissipation is responsible for the loss of the quantum coherence or suppression of topological protection.

It is the aim of our presentation to appoint to the field of quantum thermodynamics a new actor $-$ the superconducting vortex. It appears naturally in type II superconductors upon exceeding a certain magnetic field as an energetic compromise between Meissner (when magnetic field is expelled from the sample) and normal state (when magnetic field can entirely pass through the sample). Superconducting vortex is a pure quantum object: it is microscopic ring of supercurrent, which collects $2\pi$ of superconducting phase on one round trip and encircles quantized filament of magnetic flux, known as flux quantum $\Phi_0=h/2e$. As long as superconducting vortices do not move, the externally applied current $I_A$ is dissipationless, for it finds its way between vortices and preserves perfect conductivity. However, as $I_A$ is increased the Lorentz force acting on vortices may put them into motion. Mobile vortices become source of dissipation and temperature of the sample goes up owing to the creation of quasiparticles.

In our work we can trap and expel a single vortex on demand with pulses of electrical current. The supreme control over single vortex dynamics combined with the time-resolved nanothermometry\cite{Zgirski2018} allows us to measure the temperature jump after vortex has been expelled from the aluminum nanoscale sample and the subsequent thermal relaxation. We get the experimental access to the energetic cost of a single vortex expulsion. Apart from a deep insight into thermodynamics of a moving vortex we present an experimental platform for emerging field of vortex electronics\cite{Krasnov2015,Ligato2021,Steinberg2023,Kalashnikov2023}. Our device is in fact a simple memory cell, but it also shows features of a superconducting diode (fig. S1).

Our study may improve understanding of the detection mechanism of superconducting nanowire single-photon detectors\cite{Engel2015,Renema2014,Poggio2019}. It suggests that moving vortex in such devices could enhance the initial photon absorption by producing additional dissipation\cite{Bulaevskii2011} (fig. S2B).

\begin{figure}[h]
\centering
\includegraphics[width=0.9\textwidth]{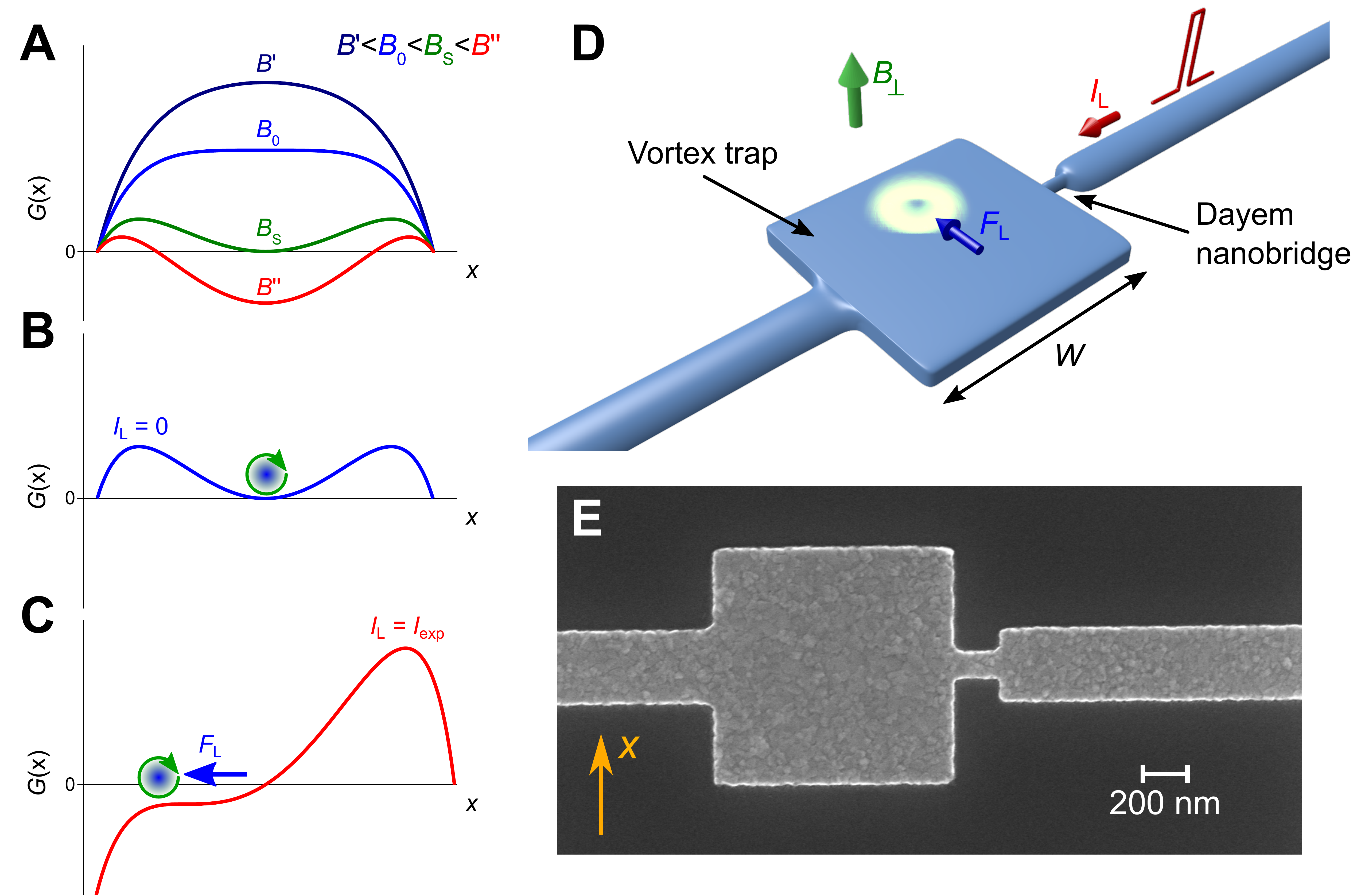}
\caption{\textbf{Single Vortex Box.} (\textbf{A}) Landscape of the Gibbs free energy of a single vortex state across the width of the box at various magnetic fields $B_{\bot}$ and with no applied current $I_L$. For $B_{\bot}>B_0$ the state with vortex becomes energetically favorable when sample is cooled across $T_c$. (\textbf{B} and \textbf{C}), The effect of the applied current on the tilt of the potential energy. For $I_L=I_{exp}$ the dependence shows no minimum that would stabilize the vortex and it leaves the sample pushed out by the Lorentz force $F_L$. $I_{exp}$ grows with the field because vortex is stronger bound in local energy minimum further away from the transition field $B_0$. (\textbf{D}), Layout of the studied nanostructure consisting of a Single Vortex Box, a Dayem nanobridge and connecting leads. The Lorentz force exerted on the vortex by the applied current $I_L$ in the presence of perpendicular magnetic field $B_{\bot}$ is depicted schematically. (\textbf{E}), SEM image of the working aluminum device.}\label{Fig_1}
\end{figure}

\section*{Theoretical background}\label{sec2}

Superconductor expels the externally applied magnetic field from its interior, owing to existence of Meissner screening currents. If the kinetic energy of these currents becomes too large it is energetically favorable for the sample to let some magnetic field lines in. Magnetic flux that enters into sample involves formation of quantized loops of supercurrent. If sample is small enough and cooled across critical temperature $T_c$ in applied magnetic field, it is possible to trap just a single vortex, provided that Gibbs free energy develops a metastable minimum (Fig.\,\ref{Fig_1}A). For superconducting strips of width $W$ such minimum is separated with the Bean-Livingston barriers from the edges of the strip and is first established when magnetic field exceeds the threshold value, i.e. $B_0=\pi\Phi_0/(4W^2)$\cite{Bean1964,Likharev1972,Maksimova1998}. This prediction holds for superconducting strips\cite{Martinis2004}, but qualitatively (up to a numerical factor of the order of unity) is also correct for the squared 0-dimensional confinements which are studied below\cite{Zgirski2023}. The presented model predicts also that the trapped vortex can be expelled from the nanostructure by application of the pulse of electrical current, which tilts the potential energy and removes the local energy minimum: owing to the Lorentz force, the vortex is pushed to the side of the square and eventually it escapes out of the sample (Fig.~\ref{Fig_1}C,D). Once the current pulse is over, the potential regains its original shape with minimum in the middle, but the vortex is not present in the nanostructure. Importantly, owing to increased depth of the potential well for larger fields, they implicate higher currents necessary to expel the vortex.

The effects related to dissipation due to moving vortices were widely studied in current-driven thin superconducting films\cite{Pannetier2005,Embon2017,Vodolazov2020}. The investigations were performed for samples containing huge number of vortices moving in steady states and the effect of dissipation was deduced from voltage appearing on the sample once the threshold value of the current bias was exceeded. Researchers identified flux flow regime\cite{Strnad1965} and avalanche regime\cite{Field1995,Altshuler2004, Yuri2017}, but the presented experiments did not give access to elementary dissipative process, which is expulsion of a single vortex from a superconductor.

\section*{Experimental approach}\label{sec3}

\begin{figure}[t]
\centering
\includegraphics[width=0.98\textwidth]{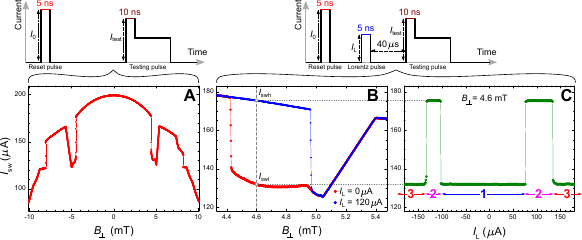}
\caption{\textbf{Electrical probing and manipulation of the vortex state.} (\textbf{A}), Switching current of the nanobridge vs. perpendicular magnetic field $I_{sw}(B_{\bot})$ reveals a pronounced dip in the characteristics for the field values where the entry of a single vortex is expected. (\textbf{B}), $I_{sw}(B_{\bot})$ dependence in the region of the dip. The red curve is a detailed measurement of the region of the suppressed switching current $I_{swl}$ visible in the curve of panel A and the blue curve presents the effect of the application of the additional pulse, called the Lorentz pulse. It is high enough to expel the vortex, but too low to switch the bridge. The following testing pulse probes the box in the Meissner state: this time there is no vortex to be expelled. Consequently, there are no quasiparticles excited in the box and the switching current remains at its high value $I_{swh}$. (\textbf{C}), Switching current of the nanobridge as a function of the Lorentz pulse amplitude recorded at the fixed magnetic field in the dip region (dashed vertical line in panel B. For low values of the Lorentz pulse (region 1), it can neither expel vortex nor switch the bridge. In region 2 Lorentz pulse can expel the vortex but not switch the bridge. Finally, for the highest values of the Lorentz pulse (region 3), it first expels the vortex and then switches the bridge. The pulse protocols used to collect the presented dependencies are displayed at the top of the figure. The bath temperature is $T_0=400\,$mK.}\label{Fig_2}
\end{figure}

We fabricate Single Vortex Box (SVB) with standard e-beam lithography by evaporating 30$\,$nm of aluminum (Fig.~\ref{Fig_1}E). It is attached to a short Dayem nanobridge, whose critical current is sensitive to the vortex state of the box\cite{Zgirski2023}. The structure (box+nanobridge) is connected to the contact pads through 15$\,\mu$m long and 300$\,$nm wide leads. Such geometry, although very simple, allows not only to monitor but also manipulate the vortices in the box with pulses of electrical current. We can initialize the box in a single vortex state (with the reset pulse), expel the vortex (with the Lorentz pulse) and detect the presence or the absence of the vortex by probing the switching current of the nanobridge with the testing pulse (see pulse protocols in Fig.~\ref{Fig_2} and text S1). Noteworthy, our approach is also compatible with time-resolved switching thermometry developed in recent years\cite{Zgirski2020} i.e. we can measure the thermal response of the trap after application of the current pulse, which changes the vortex state of the box (fig. S3-S5).

\section*{Results}\label{sec4}

We measure switching current of the nanobridge as a function of the perpendicular magnetic field $I_{sw}(B_{\bot})$ (Fig.$\,$\ref{Fig_2}A). At low values of the applied field we see continuous suppression of the critical current due to enhancement of the Meissner screening currents expelling away magnetic field lines. This regime is followed by narrow range of fields where we observe the pronounced dip in the $I_{sw}(B_{\bot})$ characteristics, being a sign of strong reduction of the superconducting order parameter due to dissipation. The abrupt transition corresponds to the first vortex penetration field $B_0\sim \frac{\pi\Phi_0}{4W^2}$. The trace is thus consistent with the interpretation in which we expel the vortex from nanostructure in the narrow field window.  In the discussed case the testing pulse of the bridge provides also the Lorentz force necessary to get rid of the vortex. The leaving vortex produces the excess population of quasiparticles, which suppresses the switching current of the bridge. We see that at higher fields $I_{sw}$ recovers to a big extent, signaling absence of the dissipative process due to the moving vortex. It happens because the current needed to expel the vortex $I_{exp}$ grows with field and becomes larger than $I_{sw}$. The $I_{sw}(B_{\bot})$ characteristics for other samples are provided in fig. S6.

It is possible to partially "heal" the dip, i.e. extend the Meissner state of the box into the higher magnetic fields by the application of the Lorentz current pulse, which expels the vortex, but does not switch the junction. The testing pulse which probes the bridge is then applied long time after expulsion, when the box comes back to thermal equilibrium, and finds the bridge in the Meissner state corresponding to a high value of the switching current $I_{swh}$. Such scenario is presented for $I_{sw}(B_{\bot})$ dependence in Fig.~\ref{Fig_2}B. Scanning the amplitude of the Lorentz pulse at a fixed magnetic field inside the cusp, one can find 3 regions (Fig.~\ref{Fig_2}C). In the first one, for the lowest values of the Lorentz pulse the switching current is also low ($I_{swl}$). Here the Lorentz pulse cannot remove the vortex, but the testing current provides enough amplitude to do it. Following the expulsion, the box warms up and as a consequence the same testing pulse probes the thermally excited state of the bridge. In the second region the Lorentz pulse expels the vortex, but it does not switch the bridge. The box is now in the Meissner state and the testing pulse finds the high value of the switching current ($I_{swh}$). Finally, in the third region, the Lorentz pulse expels the vortex but because of the too high amplitude it also switches the bridge. The bridge and box go to the normal state and in the following cooldown another vortex is trapped in the box. The Lorentz pulse works here as the second reset pulse and overall does not change the state of the box. When testing pulse arrives it finds the vortex in the box and the expulsion-switching scenario described for the region 1 follows: switching current is low again ($I_{swl}$).

\begin{figure}[h!]
\centering
\includegraphics[width=0.8\textwidth]{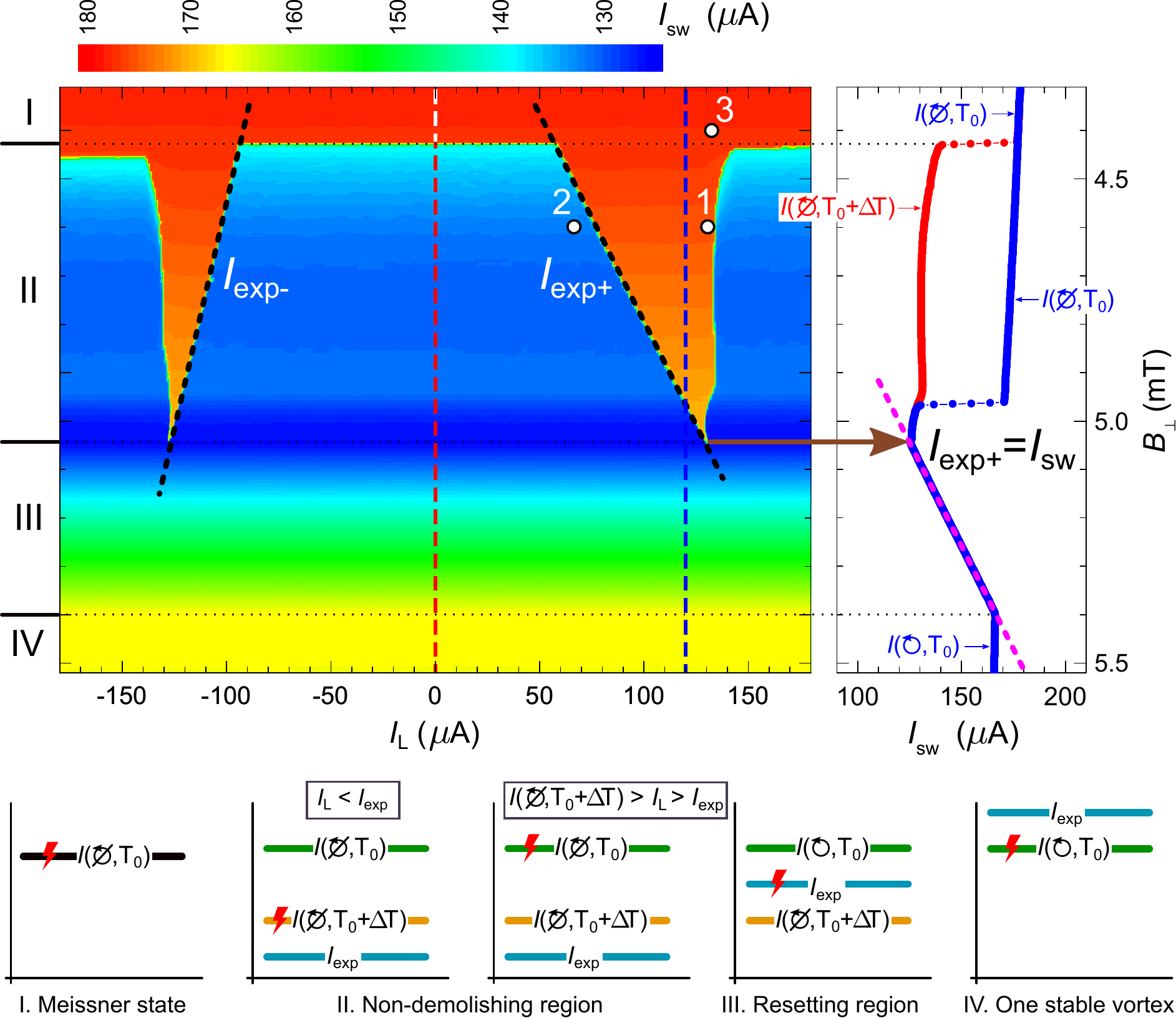}
\caption{\textbf{As-received experimental vortex stability diagram:} switching current dependence of the nanobridge on the applied magnetic field and the amplitude of the Lorentz pulse. It reveals four distinct regions in magnetic field: \textbf{I.} There is no vortex entry in the field-cooled sample after application of the reset pulse: the box remains in the Meissner state and the Lorentz pulse plays no role. \textbf{II.} Vortex is captured in the box just after the reset pulse (text S1). It can be expelled with sufficiently high Lorentz pulse without switching the bridge. The zone of the Meissner state is thus extended into higher field values, what is observed as the two triangles in the diagram. The inner slopes of triangles (indicated with dashed lines) mark the minimum value of the Lorentz pulse necessary to expel the vortex $I_{exp}(B_{\bot})$. The switching current (measured with the testing pulse) is low when it has to expel the vortex from the box, and high, if the box is in the Meissner state – see also Fig.~\ref{Fig_2}B. \textbf{III.} Vortex is captured in the box just after the reset pulse, as in region II. The required expulsion current $I_{exp}$ is higher than the switching current after vortex has been expelled i.e. application of the Lorentz pulse capable of expelling the vortex necessarily leads to the switching of the bridge, providing the reset for the box. In this region the switching current is equal to the expulsion current – see the right panel. The line $I_{exp+}(B_{\bot})=I_{sw}(B_{\bot})$ is a continuation of $I_{exp+}(B_{\bot})$ dependence from region II for larger magnetic fields ($I_L$ and $I_{sw}$ axes have the same unit revealing the same slope of the two pieces of $I_{exp+}(B_{\bot})$ relation, see fig. S1). \textbf{IV.} Vortex is captured in the box just after the reset pulse, as in region II and III. It is not possible to expel it because the required Lorentz pulse would need to be larger than the switching current of the bridge in the vortex state. The right figure contains two cross-sections of the main map which are denoted with the vertical dashed lines (the data are the same as those in Fig.~\ref{Fig_2}B). The mutual relations between switching and expulsion currents in the four regions are illustrated schematically in the bottom panel (the loop and crossed loop respectively indicate the presence or absence of a vortex in the box at the very moment when the nanobridge switches). The switching levels are schematically indicated with the lightnings.}
\label{Fig_3}
\end{figure}

The $I_{sw}(I_L)$ scan can be collected for various magnetic fields building the vortex stability diagram, i.e. the $I_{sw}(B_{\bot},I_L)$ map (Fig.~\ref{Fig_3}). The map shows regions of magnetic fields and the Lorentz current pulses where expulsion of the vortex is possible.

We focus on the zone II of the diagram (Fig.$\,$\ref{Fig_3}). Here, we can expel the vortex without switching the junction. The switching current is measured 40$\,\mu$s after application of the Lorentz pulse $I_L$. If the Lorentz pulse expels the vortex, the box has enough time to equilibrate at the bath temperature  $T_0=400\,$mK and we find high value of the switching current $I_{swh}=I$(no vortex, $T_0$). Otherwise, the testing pulse itself first expels the vortex at its rising slope (the rising time of the pulse is equal to 2-3$\,$ns, i.e. its dynamics is much slower than that of the vortex) and only then probes the thermally excited state of the box. It results in a low value of the switching current $I_{swl}=I$(no vortex, $T_0+\Delta T$). Since the switching current of the bridge is uniquely related to the temperature of the box, both $I_{swh}$ and $I_{swl}$ can be converted into temperature with the aim of the $I_{sw}(T)$ calibration (see fig. S2 and S7). The resulting temperature difference $\Delta T=T(I_{swl})-T(I_{swh})=T_h-T_l$ corresponds to the instantaneous temperature increase due to the expulsion of the vortex out of the box. For the presented map $T_l=T_0=400\,$mK and $T_h\sim 650\,$mK. Defining the volume of the SVB as $\Omega_B=ta^2$ (with $t=30\,$nm and $a=1\,\mu$m denoting the thickness of the superconducting box and the length of its side, respectively), and taking the dependence of the aluminum heat capacity $C_p(T)$ from literature\cite{Philips1959} as a reasonable approximation, the measured $\Delta T$ yields

\[  \Delta Q = \Omega_B \int_{0.4\,K}^{0.65\,K} C_p(T) \,dT = 4.3\cdot 10^{-19}\,J\]

as a calorimetric estimation of the released heat. This energy is equivalent to the absorption of a single photon of the red light. Importantly, in the studied range of parameters we do not observe any significant changes of this energy with magnetic field $B_{\bot}$ or expulsion current $I_{exp}$: the measured temperature rise after the vortex expulsion $\Delta T$ remains very similar as it is evident from almost constant values of $I_{swh}$ and $I_{swl}$ visible as the red and blue regions in Fig.$\,$\ref{Fig_3}.

In the picture of the viscous flow of the vortex, the dissipation of energy in the superconducting box occurs as the vortex sweeps through a specific volume, causing the transformation of Cooper pairs into quasiparticles. The volume within which the conversion occurs can be defined by the trajectory of the moving vortex, i.e. as $\Omega_v = \frac{1}{2}at\xi$, where $\xi\cong150\,$nm is an estimate for the coherence length of the superconductor at 400$\,$mK. Since the superconducting pairing involves only electrons at the surface of the Fermi sea, the number of quasiparticles created in the process is $n_{qp}=g(E_f) \Omega_v \Delta$, where $g(E_f)$ is the density of states at the Fermi level and $\Delta=1.76k_BT_c=200\,\mu$eV is the superconducting gap. The required excitation energy is roughly $\Delta Q_{v}=n_{qp}\Delta=3.1\cdot10^{-19}\,$J and matches the magnitude of the dissipated heat $\Delta Q$ found in the experiment (see text S2 for a complementary discussion).

\begin{figure}[h]%
\centering
\includegraphics[width=0.9\textwidth]{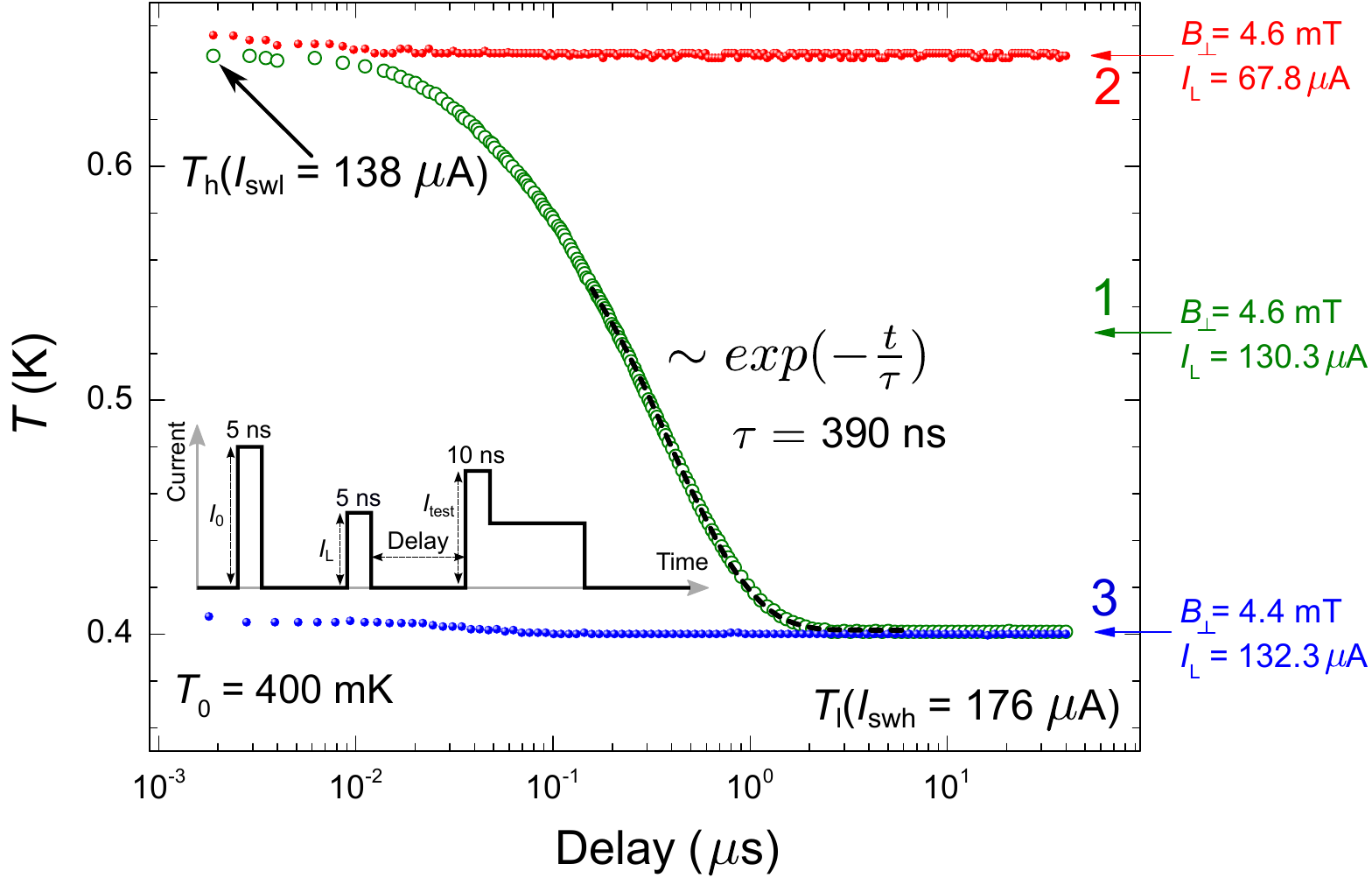}
\caption{\textbf{Experimental thermal dynamics of the SVB after expulsion of a single vortex} with the $I_L=130.3\,\mu$A at $B_{\bot}=4.6\,$mT (curve 1). The other two curves are references revealing no temperature variation after the application of either a too low Lorentz pulse to expel the vortex ($I_L=67.8\,\mu$A, curve 2) or a too low magnetic field to trap the vortex after the application of the reset pulse ($B_{\bot}=4.4\,$mT, curve 3). In the case of the curve 2 the testing pulse itself expels the vortex. It results in the elevated value of the probed temperature, which is independent of the delay. The broken line, imposed on the curve 1, represents the exponential fit in the linear regime. The three points ($I_L,B_{\bot}$), corresponding to the three curves, are imposed on the vortex stability diagram in Fig.\,\ref{Fig_3}.}\label{Fig_4}
\end{figure}

The excited quasiparticles spread immediately in the SVB - the diffusion time across the box is of the order of 100\,ps, and equilibrate with other electrons, which leads to the increase in the thermodynamic temperature of the box. Using the protocol of the nanosecond-resolving switching thermometry\cite{Zgirski2018}, we can measure the temporal relaxation profile of the box after expulsion of the vortex. It is accomplished by measuring the switching current of the bridge for various delays between the Lorentz and testing pulse (Fig.$\,$\ref{Fig_4}). The relaxation time in linear regime is 390$\,$ns, in agreement with the thermal relaxation times for the aluminum nanowires studied by us in the earlier works, where either the switching of the nanobridge to the normal state\cite{Zgirski2018} or the Joule heating of metallic island\cite{Zgirski2020} were used to excite the quasiparticles. The time is slightly smaller than expected from the electron-phonon relaxation channel alone due to the significant role of the quasiparticle diffusion along the leads and non-zero value of the magnetic field. The flat profile of the relaxation curve during first 10$\,$ns, where relaxation has hardly started, allows us however to neglect the hot electron diffusion in the estimation of the dissipated energy $\Delta Q$. Systematic studies of thermal relaxations triggered by expulsion of the vortex and measured for various $B_{\bot}$ and $I_L$ are presented in fig.~S3-S5.

\section*{Analysis/Discussion}\label{sec5}

Our experiment allows to trigger dissipation on demand in the self-limiting elementary process: we have only one vortex which we can expel. One may envisage the demonstrated scheme as a convenient way to generate a limited number of quasiparticles in a superconducting structure. It is expected that the dynamics of the expulsion process is of the order of a few tens of picoseconds. The expulsion itself is too fast to be observed experimentally but it produces the measurable thermal trace. We refer to the vortex expulsion as the 'delta-heating' because it offers a triggerable  and extremely narrow time frame during which a definite number of quasiparticles is generated. This is in stark contrast to phase slips, often referred to as 1D vortices, which are indirectly observed in superconducting junctions and 1D nanowires near the critical current. Phase slips represent stochastic dissipative events that eventually result in the transition of the samples to the normal state \cite{Bezryadin2009,Zgirski2021}.

Similarly to the phase slip process, conveniently considered in the landscape of the tilted-washboard potential\cite{Tinkham2004}, the vortex expulsion may involve thermally stimulated jump over the Bean-Livingston energy barrier (Fig.~\ref{Fig_1}). Both processes are stochastic in the narrow range of electric currents. The transition between the blue region (vortex not expelled with the Lorentz pulse) and the red region (vortex expelled with the Lorentz pulse) in zone II of the vortex stability diagram (Fig.~\ref{Fig_3}), as seen along the $I_{exp}(B_{\bot})$ line, is very sharp. However, by using a finer resolution for the Lorentz pulse, we can accurately measure the current-dependent probability of the vortex escape (fig.~S8). Alternatively to the thermal activation, the vortex expulsion may possibly proceed through macroscopic quantum tunneling, a phenomenon widely studied not only in superconducting junctions\cite{Martinis1988} and wires\cite{Zgirski2005}, but also in magnetic clusters\cite{Barbara1997}.

\section*{Conclusion}\label{sec6}

We demonstrate operation of a Single Vortex Box (SVB), in which vortex can be manipulated similarly to an electron in a Single Electron Box\cite{Grabert1992}. Thus, vortex can be treated as a macroscopic, albeit quantized "particle", which can be created and annihilated with pulses of electrical current. This feature combined with the fast time-resolving thermometry provides a comprehensive experimental insight into the physics of moving vortices.

The expulsion of a single superconducting vortex with the current pulse from a mesoscopic sample produces dissipation at the level of 4$\cdot$10$^{-19}\,$J. This is the energy necessary to turn all Cooper pairs into quasiparticles on the path of the escaping vortex.

Our study provides an example of a fundamental intrinsic dissipation in a superconducting device. Unlike a dissipation arising from fluctuations in 1D superconducting wires, where thermal or quantum phase slips play a role, the vortex-driven dissipation reported in our work can be triggered with the current pulse on demand, enabling the fully predictable rise in the temperature of a superconductor without exceeding $T_c$.

The supreme level of control of the vortex state and small size makes the presented SVB an attractive device for memory and logical applications in the field of superconducting/vortex electronics. In such realization three pulses in the presented experiment are responsible for initialization of the memory cell, write operation (executed with the Lorentz pulse) and read-out (performed with the testing pulse). The SVB exhibits also the diode effect visible in the the vortex stability diagram. Our experimental platform is well-suited for verifying the possibility of adiabatic manipulation of vortices, which is necessary to operate them as true quantum objects.

\newpage

\topmargin -1.5cm
\oddsidemargin 0.0cm
\textwidth 17cm 
\textheight 23cm
\footskip 0.5cm

\section*{Supplementary Materials for "Quantum thermodynamics with a single superconducting vortex"} 


\baselineskip24pt

\makeatletter
\renewcommand{\thefigure}{S\arabic{figure}}
\makeatother

\subsection*{Text S1\hspace{1.5em}Pulse protocol for probing and manipulating a single vortex and measuring electron temperature of the SVB}

We perform our experiment in the dilution refrigerator. For aluminum bridges presented in this work their switching current is very close to the critical current defining transition from the superconducting to the normal state. The method is based on testing of the bridge with train of $N$ identical current pulses\cite{Zgirski2015}. The switching current $I_{sw}$ is defined as the one for which the switching probability of the bridge $P$ is equal 0.5. The testing pulses are repeated with period of 200$\,\mu$s guaranteeing a complete thermalization of the sample after each pulse. The switching current of the bridge is sensitive to the local population of quasiparticles, which in turn depends both on temperature and distribution of the Meissner screening currents. The first property makes the bridge a sensitive thermometer\cite{Zgirski2018}, and the second one allows for the detection of magnetic field and vortices, even if they are not expelled\cite{Zgirski2023}. The standard probing protocol is extended by the application of two additional pulses in each cycle, which precede the actual testing pulse (Fig. 2, pulses). The first prepulse is so called reset pulse $I_0$, for its amplitude is significantly higher than the switching threshold of the bridge. Its role is to transit sample to the normal state and overheat the vortex box above $T_c=1.3\,$K. The subsequent cooling, taking place in the presence of applied magnetic field allows to trap the vortex inside box, initializing the sample in a well-defined state. It takes around $20\,$ns for our structure to cool down back to $T_c$ once the reset pulse is switched off \cite{Zgirski2018}. The reset pulse can be thus also thought of as a trapping pulse or initializing pulse. The second prepulse is intended to change the vortex state of the box without switching the bridge. It is called the Lorentz pulse $I_L$, owing its name to the force it exerts on the vortex. Setting the time delay between the Lorentz and testing pulses one can measure the temporal relaxation of the switching current of the bridge, which is a thermal consequence of the dissipative dynamics of a single vortex. For fast electron-electron interaction the presence of excess quasiparticles is thermodynamically equivalent to elevation of their temperature. This allows us to convert the measured dependence of the switching current into the temporal profile of the electron temperature in the SVB, which follows the expulsion of the vortex from the trap.

\subsection*{Text S2\hspace{1.5em}Analysis of the dissipated energy}

The dissipated heat comes from the Gibbs free energy difference between vortex and Meissner states $\Delta G=G_{v}-G_{M}$ and the work $W$ done by the current source to expel the vortex. We can calculate the energy delivered from the current source during expulsion of the vortex with current $I_{exp}$ and voltage $V$ across the box by integrating instantaneous power over time window $\tau$ when vortex is being expelled:
\[  \Delta E_s(B_{\bot}) = \int_{0}^{\tau} VI_{exp} \,dt\]

Using 2$^{nd}$ Josephson relation we get:

\[  \Delta E_s(B_{\bot}) = \int_{0}^{\tau} \frac{d\varphi}{dt}\frac{\Phi_0}{2\pi}I_{exp} \,dt=\frac{\Phi_0I_{exp}}{2\pi} \int_{0}^{\pi} \,d\varphi=\frac{\Phi_0I_{exp}}{2}\]

It is analogous formula to that for the energy dissipated in the phase slip event of the Josephson junction if to replace  the critical current of the junction $I_c$ with $I_{exp}$ and notice that vortex leaving the box is equivalent to a half of phase slip, i.e. the phase across the box changes by value close to $\pi$. This energy ranges from $E_{s}=6\cdot 10^{-20}\,$J to $E_{s}=1.2\cdot 10^{-19}\,$J for $I_{exp}=60\,\mu$A to $I_{exp}=120\,\mu$A (cf. Fig. 3), which are the values 5 to 2.5 times smaller than the amount of the measured dissipated energy $\Delta Q$. It suggests that in the observed process the significant fraction of the dissipated energy comes from the Gibbs free energy difference between vortex and Meissner state of the box $\Delta G=G_{v}-G_{M}$, i.e. the vortex state corresponds to the local energy minimum, but is not absolutely stable, as schematically presented in Fig. 1A for field $B_{\bot}$ higher than $B_0$ but smaller than $B_s$. Such interpretation would support claim\cite{Clem1998,Maksimova1998,Martinis2004} that in the cooled-down superconducting nanowire magnetic field lines are first trapped when the condition for metastable equilibrium is met. It is noteworthy that the initial (vortex) and final (Meissner) state of the box are obviously physically different. It is in contrast to the phase slip process in a junction or nanowire, in which the initial and final states are the same.

When increasing magnetic field the state with vortex becomes energetically more favorable (the local energetic minimum becomes deeper), i.e. we get less heat from the free energy when transferring the SVB from the vortex to Meissner state (in fact $\Delta G$ may turn negative at higher magnetic fields than these studied in the presented experiment). At the same time we start to dissipate more energy from the current source. The energetic balance reads: $\Delta Q=W+\Delta G$. In the studied range of parameters, we do not observe any significant variation of $\Delta Q$ with $B_{\bot}$ (see Fig.~S3). Moreover, once the vortex is expelled, we also see no difference in the amount of the dissipated energy $\Delta Q$ for the two polarities of the Lorentz pulse, although there is a significant difference in the expulsion threshold for them (see Fig. S9).

\newpage

\begin{figure}[h]
\centering
\includegraphics[width=0.7\textwidth]{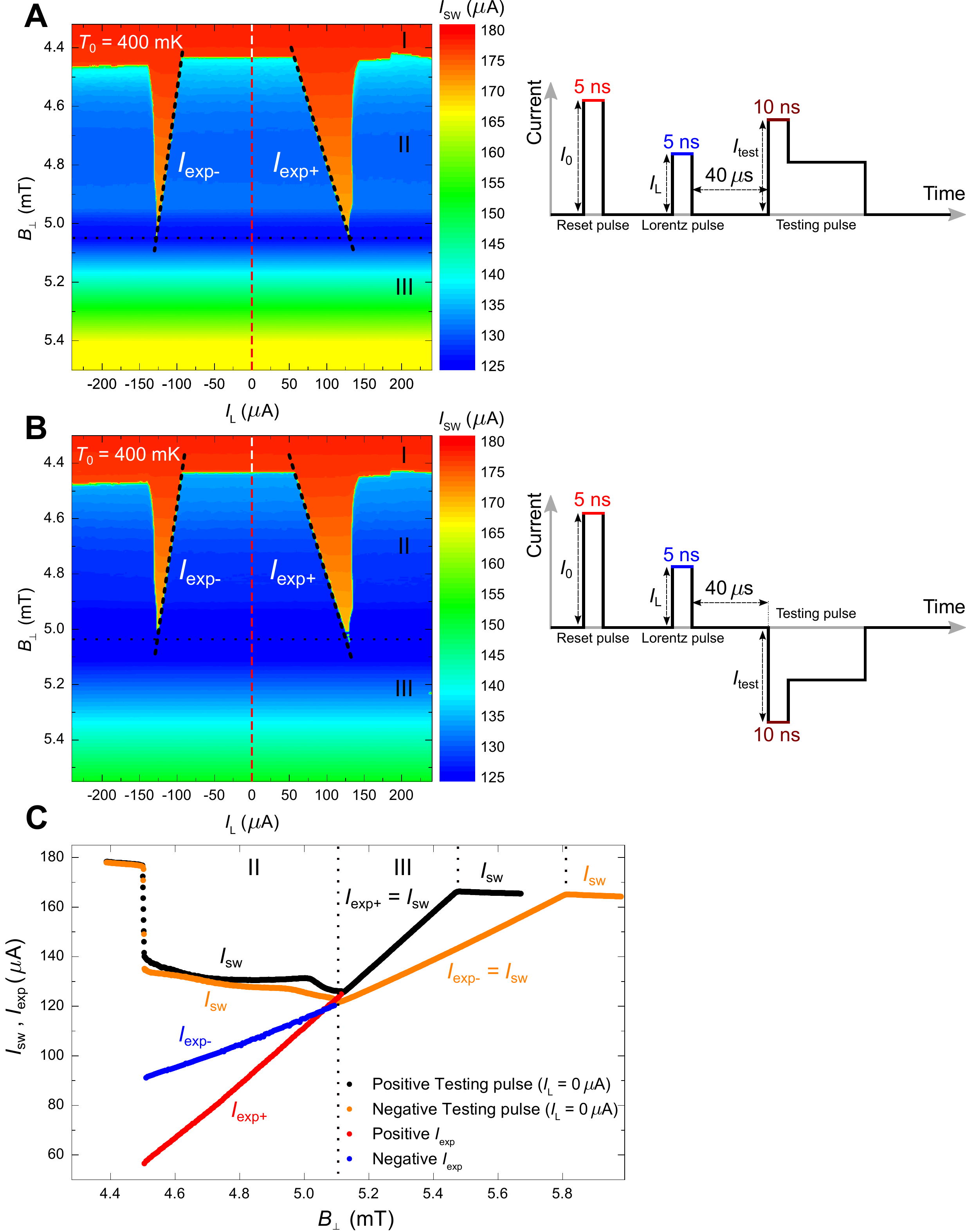}
\label{Fig_EC}
\end{figure}
\noindent Figure S1: \textbf{Vortex diode effect.} (\textbf{A}) Vortex stability diagram collected for the positive testing pulses. (\textbf{B}) Vortex stability diagram collected for the negative testing pulses. The broken lines $I_L=0$ and $I_L=I_{exp}$ indicate curves presented in panel C. (\textbf{C}) The switching current and the expulsion current dependencies on magnetic field.  The two non-monotonous curves are the indicated cross-sections of the vortex diagrams at $I_L=0$ collected for the positive and negative testing currents. The red and blue lines are the inner edges of the vortex stability diagrams which correspond to the onset of the vortex expulsion with the positive and negative Lorentz pulses $I_L$ respectively (note that the positions of the inner edges does not depend on the polarity of the testing pulse, as expected). Importantly, the two curves for each polarity match together when crossing from zone II to III of the diagram creating the two non-symmetric $I_{exp}(B_{\bot})$ almost linear dependencies for the two polarities. This evidences the diode effect. In the zone II it is the Lorentz pulse that expels the vortex (the switching current of the nanobridge is larger than the expulsion current, i.e. it is possible to expel the vortex without switching the bridge), in the zone III the testing pulse itself expels the vortex which leads to the switching of the junction (the switching current of the nanobridge after expelling the vortex is smaller than the expulsion current, i.e. expulsion of the vortex necessarily leads to the switching of the bridge).

\newpage

\begin{figure}[h]
\centering
\includegraphics[width=0.9\textwidth]{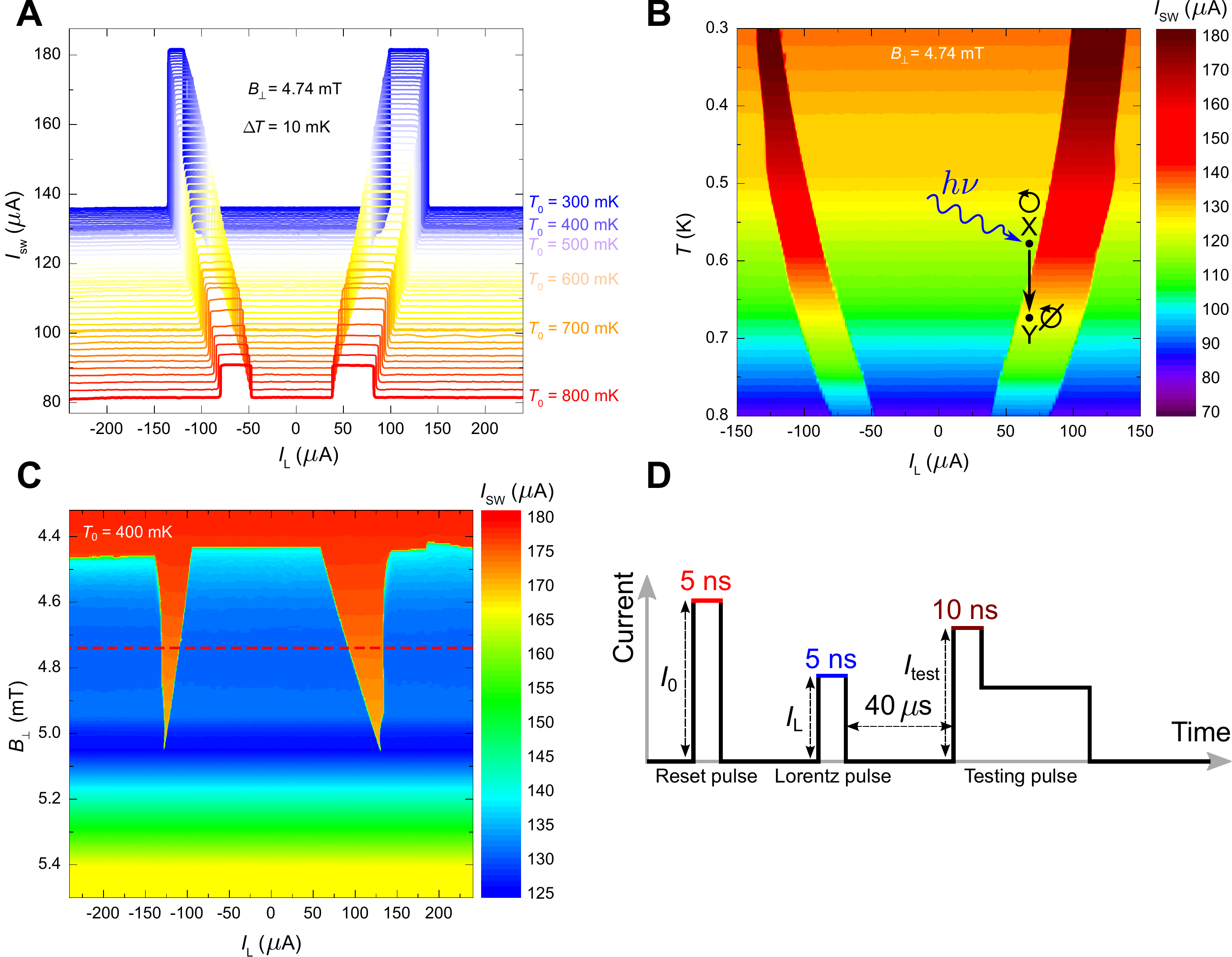}
\label{Fig_EA}
\end{figure}
\noindent Figure S2: \textbf{Temperature dependence of the vortex expulsion current and the nanobridge switching current.} (\textbf{A}) Switching current of the nanobridge vs. the Lorentz pulse amplitude recorded for various temperatures at fixed magnetic field. Inner edges of the high switching current plateaus mark the temperature dependent vortex expulsion current  for the two polarities. Outer edges are the nanobridge switching thresholds. (\textbf{B}) The same data as in A presented as a contour plot $I_{sw}(T, I_L)$ illustrating the principle of the vortex amplifier for the enhanced photon detection. The vortex box biased with $I_L$ just below the expulsion threshold, when heated by photon, will experience temperature rise (see the shift from point X to Y in panel B across the vortex expulsion edge). In point Y vortex is not stable any more. The dissipation assisting the expulsion provides much higher increase in temperature of the box than that initially caused by the photon absorption, and facilitates the detection process. (\textbf{C}) Vortex stability diagram from the main text recorded at 400$\,$mK. The dashed line marks the one chosen $I_{sw}(I_L)$ dependence (for $B_{\bot}=4.74\,$mT) which can be also found in panels A and B (for $T_0=400\,$mK). (\textbf{D}) Testing protocol: for each Lorentz pulse amplitude $I_L$ the presented sequence is repeated 1000 times to measure the switching probability of the nanobridge $P$. The amplitude of the testing pulse is adjusted with bisection algorithm to find the switching current defined as the one for which $P=0.5$. For a given temperature (panels A and B) the low value of the switching current means the vortex expulsion on the rising slope of the testing pulse, and the high value of the switching current indicates the vortex expulsion during application of the Lorentz pulse.

\newpage

\begin{figure}[h]
\centering
\includegraphics[width=0.8\textwidth]{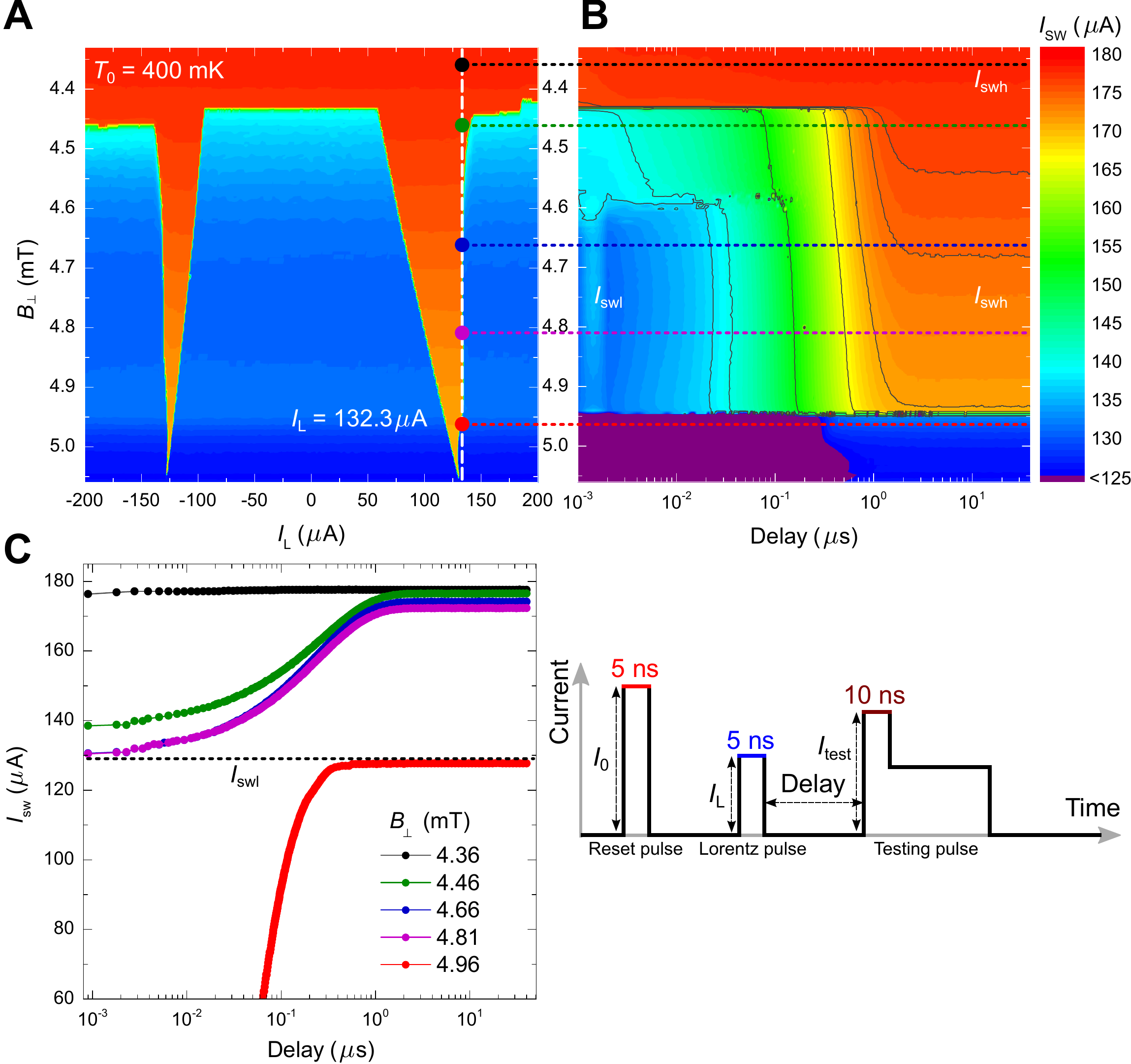}
\label{Fig_ED}
\end{figure}

\noindent Figure S3: \textbf{Thermal relaxation after expulsion of the vortex $-$ magnetic field study.} (\textbf{A}) Vortex stability diagram. Vertical dashed line shows the range of magnetic field for the relaxation experiment presented in panel b ($I_L=132.3\,\mu$A). (\textbf{B}) Thermal relaxation map $I_{sw}(B, delay)$ collected for $I_L=132.3\,\mu$A. Five horizontal dashed lines correspond to cross-sections of the map displayed in panel C. (\textbf{C}) Relaxation profiles for chosen values of magnetic field. We can distinguish 3 qualitatively different regions:

1. At low field (4.36$\,$mT), vortex is absent in the structure and therefore the Lorentz pulse does not induce dissipation: we get the flat response.

2. Above the entry field, vortex can be expelled out of the structure without switching the junction. We observe the thermal relaxation of the switching current arising from the single vortex expulsion. The difference between the two branches of the relaxation curves ($B_{\bot}=4.46\,$mT and $B_{\bot}=4.66\,$mT) may arise from not perfect fidelity in the initialization of the vortex state for smaller values of the magnetic field. As a result the relaxation curves recorded in this region ($B_{\bot}=4.46\,$mT) are the weighted averages of curves which would be measured with perfect fidelity and flat response (like that collected for $B_{\bot}=4.36$ mT).

3. For too high magnetic field ($>4.96\,$mT), the Lorentz pulse not only expels the vortex but also necessarily leads to the switching of the junction. The structure is heated above $T_c$ and once the Lorentz pulse is over, the box starts to cool-down trapping another vortex. When reading-out the bridge with testing pulse we have to expel the vortex, which leads to the additional dissipation: the thermal relaxation has an asymptote equal to the $I_{swl}$ of the vortex stability diagram (see horizontal dashed line).

\newpage

\begin{figure}[h]
\centering
\includegraphics[width=0.85\textwidth]{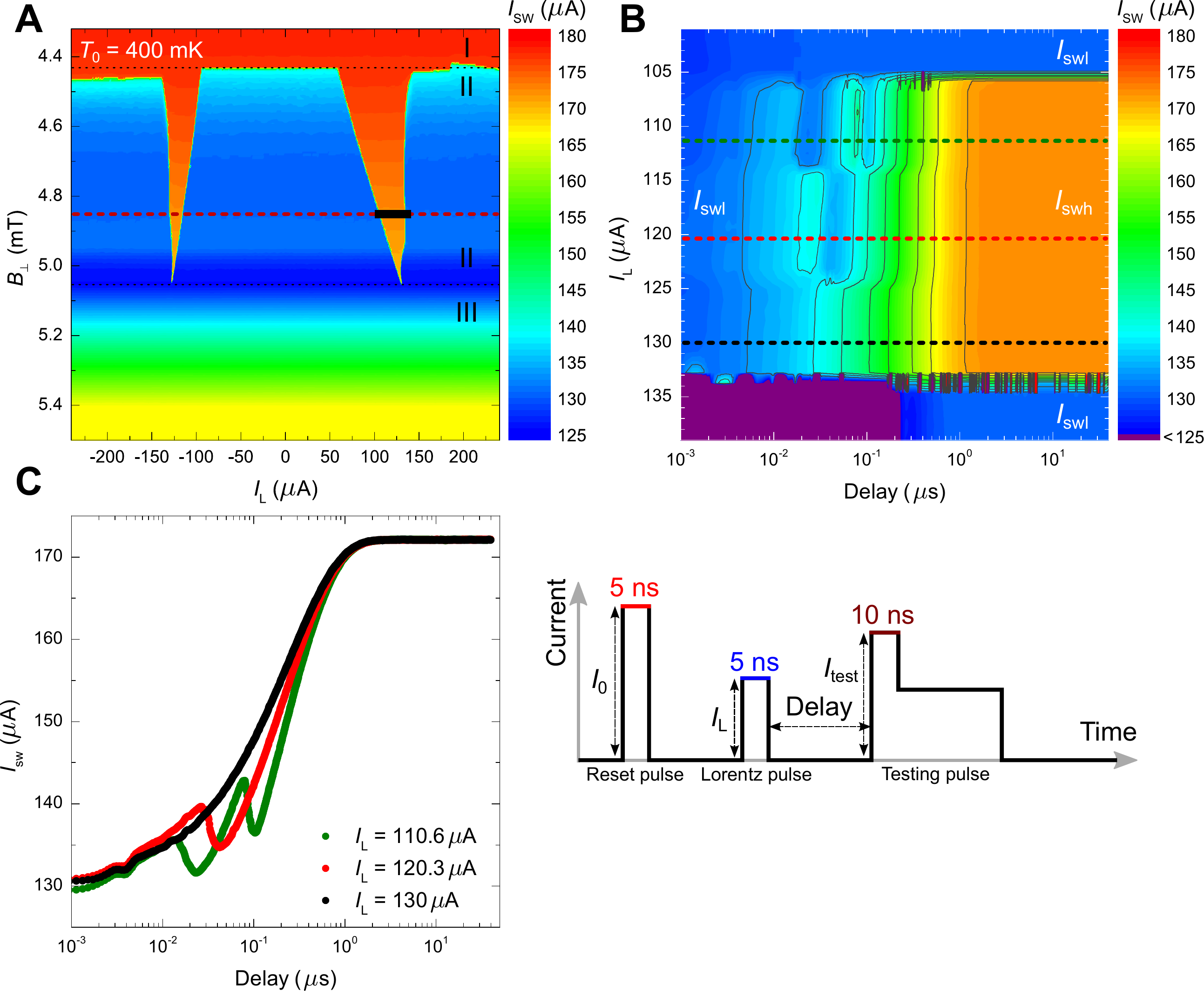}
\label{Fig_EE}
\end{figure}

\noindent Figure S4: \textbf{Thermal relaxation after expulsion of the vortex $-$ Lorentz pulse study (part I).} (\textbf{A}) Vortex stability diagram. Horizontal  black thick line shows the range of the used Lorentz pulse amplitudes for the relaxation experiment presented in panel b ($B_{\bot}=4.855\,$mT). (\textbf{B}) Set of relaxation curves $I_{sw}(delay)$ collected for various $I_L$ at fixed magnetic field $B_{\bot}=4.855\,$mT. We can distinguish three qualitatively different regions:

1. For $I_L<105\,\mu$A we see no effect of the Lorentz pulse. It is too low to push out the vortex from the box. Testing pulse expels the vortex - it leads to heating, and then switches the junction. We observe low value of the switching current $I_{swl}$.

2. In the region $105\,\mu$A$\,<I_L<133\,\mu$A vortex is expelled with the Lorentz pulse without switching the junction. We can measure the subsequent thermal relaxation with the testing pulse by varying its delay. The three horizontal dashed lines correspond to the relaxations displayed in panel C.

3. For $I_L>133\,\mu$A the Lorentz pulse not only expels the vortex but also switches the junction. We can measure the subsequent thermal relaxation from the normal state with the testing pulse. Similarly to the red curve presented in Fig. S3C each time we test the bridge, we expel the vortex introducing another heat into the nanostructure. As a result the relaxation curve is shifted towards smaller values of the switching current. Its asymptote is equal to $I_{swl}$, indicating the switching process which immediately follows the expulsion of the vortex i.e. the vortex expulsion happens on the rising slope of the testing pulse.

(\textbf{C}) Relaxation profiles for the chosen values of the Lorentz pulse. All three curves show the same relaxation time. The profiles measured at the lower Lorentz pulse amplitudes show a non-monotonous behaviour featuring one or two peaks. Only data measured for the Lorentz pulses whose amplitude is close to the switching threshold display a monotonous relaxation. We observe a sharp transition in the appearance of the relaxation curves as we increase the amplitude of the Lorentz pulse. Although not fully understood, we associate the non-monotonous relaxations with the dynamic trapping of quasiparticles in Andreev bound states of the nanobridge: when the nanostructure is cooled down some of superconducting channels in the bridge are blocked [such blocking effect is referred to as "poisoning" in the literature\cite{Zgirski2011,Siddiqi2014}], which results in the suppression of the critical current of the nanobridge.

\newpage

\begin{figure}[h]
\centering
\includegraphics[width=0.85\textwidth]{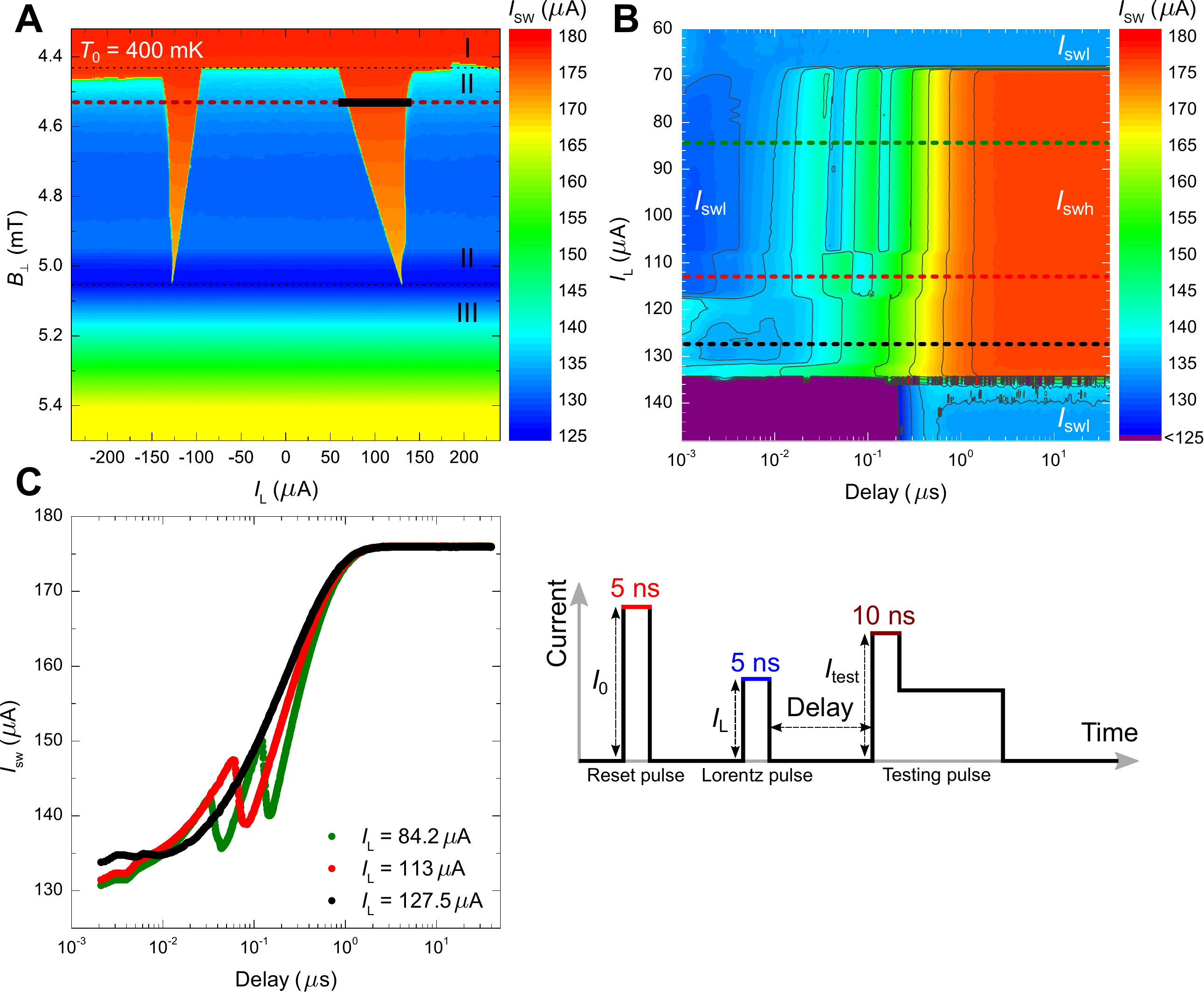}
\label{Fig_EF}
\end{figure}

\noindent Figure S5: \textbf{Thermal relaxation after expulsion of the vortex $-$ Lorentz pulse study (part II).} (\textbf{A}) Vortex stability diagram. Horizontal  black thick line shows the range of the used Lorentz pulse amplitudes for the relaxation experiment presented in panel B ($B_{\bot}=4.523\,$mT). (\textbf{B}) Set of relaxation curves $I_{sw}(delay)$ collected for various $I_L$ at fixed magnetic field $B_{\bot}=4.523\,$mT. We can distinguish three qualitatively different regions:

1. For $I_L<67\,\mu$A we see no effect of the Lorentz pulse. It is too low to push out the vortex from the box. The testing pulse expels the vortex - it leads to the heating, and then switches the junction. We observe low value of the switching current $I_{swl}$.

2. In the region $67\,\mu$A$\,<I_L<133\,\mu$A vortex is expelled with the Lorentz pulse without switching the junction. We can measure the subsequent thermal relaxation with the testing pulse by varying its delay. The three horizontal dashed lines correspond to the relaxations displayed in panel C.

3. For $I_L>133\,\mu$A the Lorentz pulse not only expels the vortex but also switches the junction. We can measure the subsequent thermal relaxation from the normal state with the testing pulse, but each time we test the bridge we expel the vortex introducing another heat into the nanostructure. As a result the relaxation curves are shifted towards smaller values of the switching current. Their asymptote is equal to $I_{swl}$, indicating the switching process which immediately follows the expulsion of the vortex i.e. the vortex expulsion happens on the rising slope of the testing pulse.

(\textbf{C}) Relaxation profiles for the chosen values of the Lorentz pulse. All three curves show the same relaxation time. The profiles measured at the lower Lorentz pulse amplitudes show a non-monotonous behaviour featuring one or two peaks. Only data measured for the Lorentz pulses whose amplitude is close to the switching threshold display a monotonous relaxation. We observe a sharp transition in the appearance of the relaxation curves as we increase the amplitude of the Lorentz pulse. Although not fully understood, we associate the non-monotonous relaxations with the dynamic trapping of quasiparticles in Andreev bound states of the nanobridge: when the nanostructure is cooled-down some of the superconducting channels in the bridge are blocked [such blocking effect is referred to as "poisoning" in the literature\cite{Zgirski2011,Siddiqi2014}], which results in the suppression of the critical current of the nanobridge.

\newpage

\begin{figure}[h]
\centering
\includegraphics[width=0.85\textwidth]{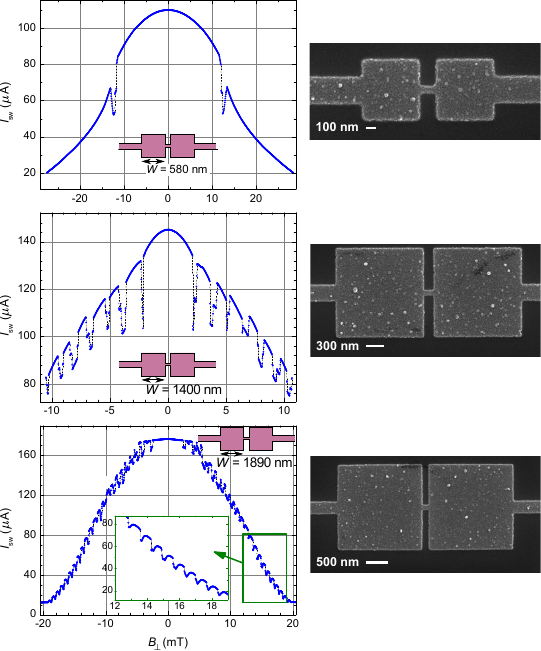}
\label{Fig_EJ}
\end{figure}

\noindent Figure S6: \textbf{Switching current of the nanobridge vs. perpendicular magnetic field $B_{\bot}$ for various vortex traps ($T_0=400\,$mK).} Samples reveal drops in $I_{sw}(B_{\bot})$ for field values where the entries of successive vortices are expected. The suppressed switching currents are measured right after expulsion of the vortex (or 2 vortices). The vortex is expelled on the rising slope of the testing pulse - it breaks Cooper pairs (rises temperature), which lowers the switching current of the nanobridge. The field for first vortex entry is well predicted with eq. (1) in Ref.$\,$\cite{Zgirski2023}. In between the dips vortices can not be expelled with the current: the expulsion current scales with $B_{\bot}$ and becomes larger than the switching current. The majority of dips for sample $W=1400\,$nm show two levels of the suppression: the deeper one corresponds to simultaneous expulsion of the two vortices, one from each trap; the shallower one is due to the expulsion of a single vortex from one trap when the vortices in the second trap become stable and can not be moved out by the current pulse. At higher magnetic fields the sample with $W=1890\,$nm shows very regular $I_{sw}$ oscillations (see inset). The geometry and morphology of the traps is provided in SEM images.

\newpage

\begin{figure}[h]
\centering
\includegraphics[width=0.8\textwidth]{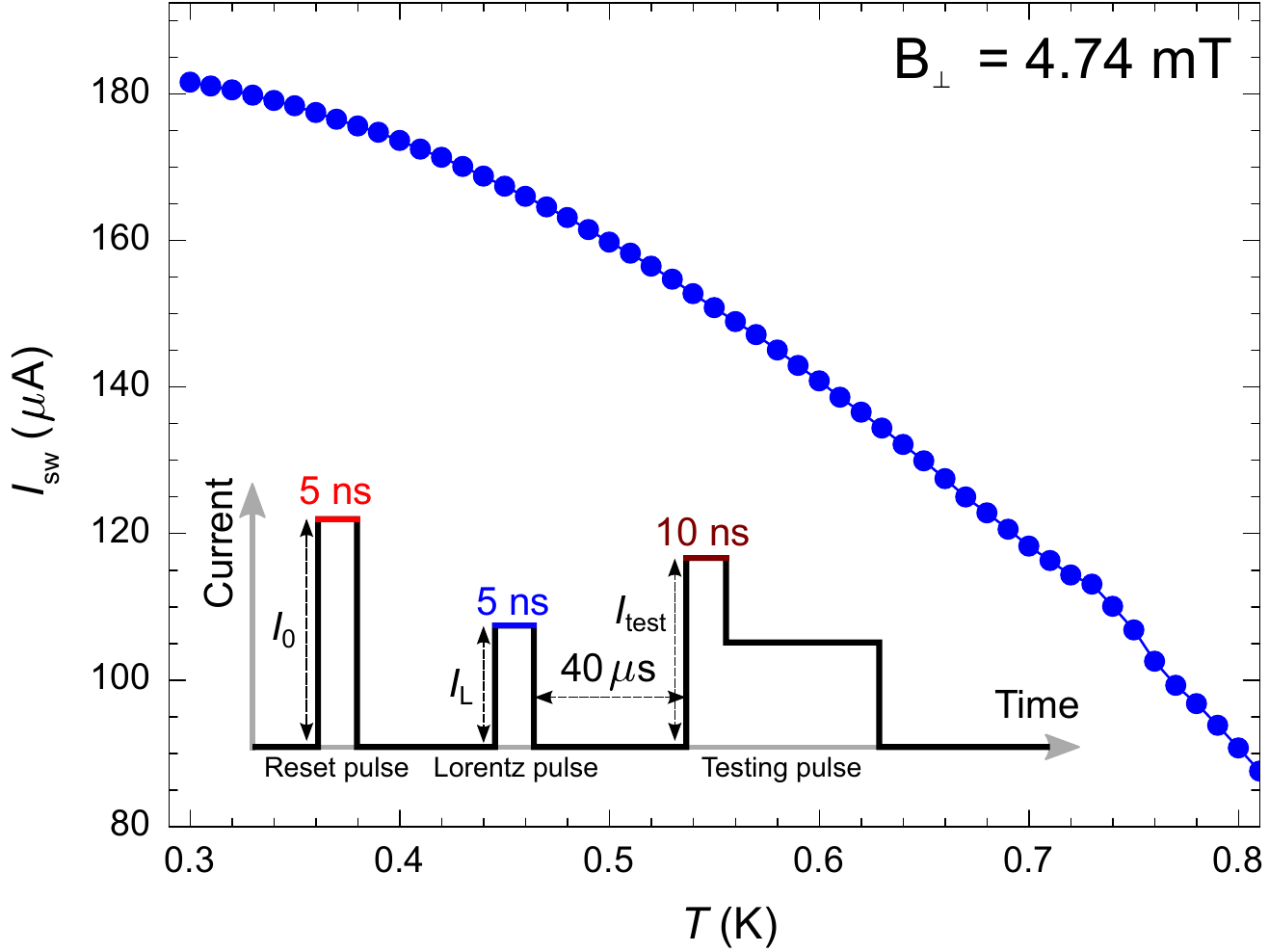}
\label{Fig_EB}
\end{figure}

\noindent Figure S7: \textbf{Calibration curve $I_{sw}(T)$ in the state with no vortex.} Switching current of the nanobridge $I_{sw}$ vs. the bath temperature $T_0$ measured after expulsion of the vortex in the thermally relaxed state for $B=4.74$ mT (zone II). The values are the plateaus visible in $I_{sw}(I_L)$ curves presented in panels A and B of the Fig. S2. The calibration curve is used to recalculate the measured $I_{sw}(delay)$ profiles into temporal dynamics of temperature $T(delay)$ after expulsion of the vortex.

\newpage

\begin{figure}[h]
\centering
\includegraphics[width=0.85\textwidth]{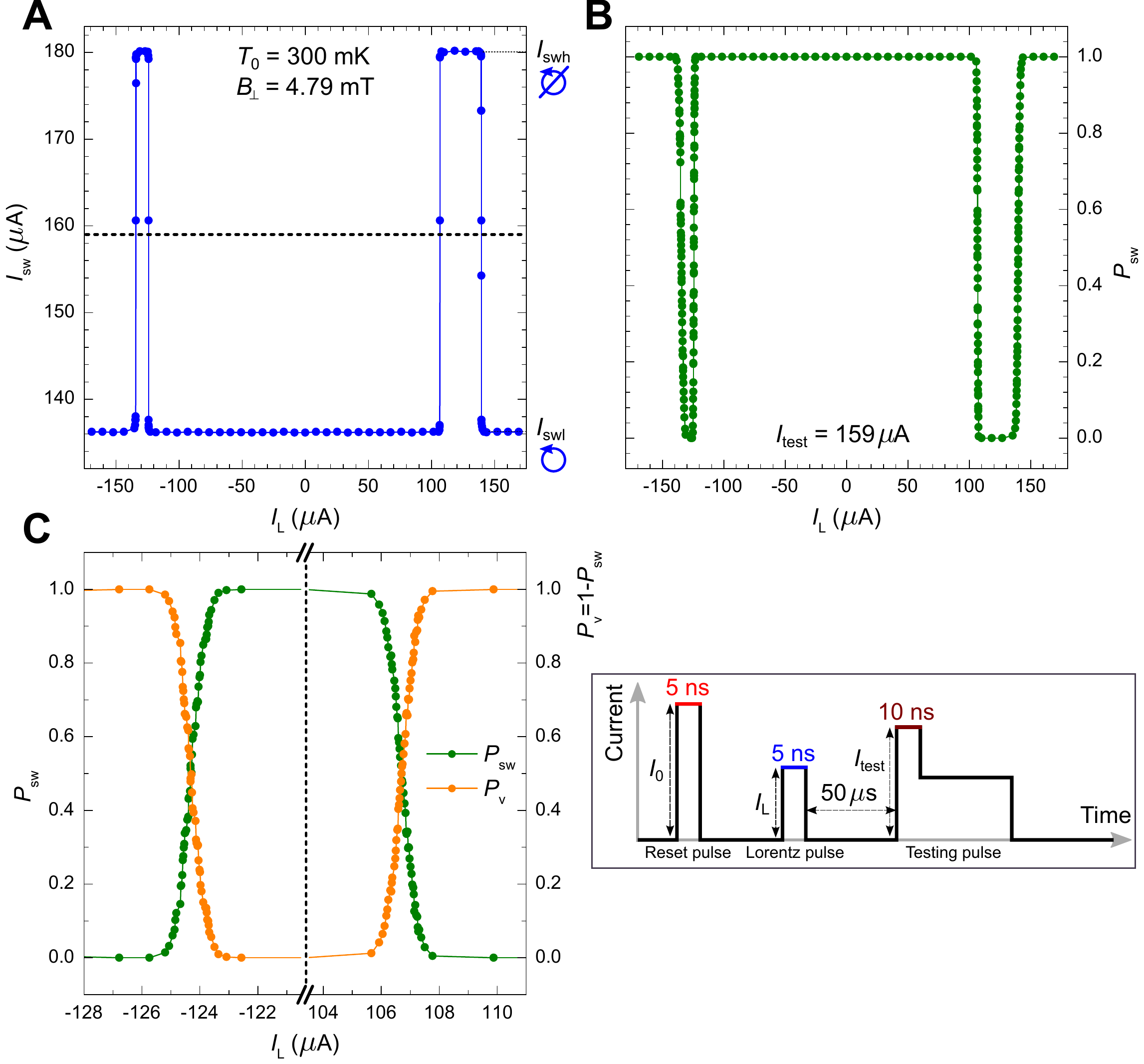}
\label{Fig_EH}
\end{figure}

\noindent Figure S8: \textbf{Vortex expulsion probability.} (\textbf{A}) Switching current of the nanobridge measured with the testing pulse recorded as a function of the Lorentz pulse amplitude at constant magnetic field. $I_{swl}$ and $I_{swh}$ correspond to the presence and absence of the vortex in the box during read-out, respectively.
(\textbf{B}) Switching probability $P_{sw}$ of the nanobridge measured as a function of the Lorentz pulse amplitude for the fixed testing current pulse $I_{test}=159\,\mu$A at the constant magnetic field $B_{\bot}=4.79\,$mT. $I_{test}$ is equal to $(I_{swl}+I_{swh})/2$ (see horizontal dashed line in panel A). The bridge necessarily switches during the testing pulse if the vortex is present in the box (i.e. it has not been expelled by the Lorentz pulse or has been trapped there upon cooling-down from the normal state after application of the too high Lorentz pulse), but it never switches if the vortex is absent (i.e. it has been expelled by the Lorentz pulse). (\textbf{C}) Magnified regions of the $P_{sw}(I_{L})$ curve presented in panel B (left axis) and the resulting vortex expulsion probability $P_{v}=1-P_{sw}$(right axis).

\newpage

\begin{figure}[h]
\centering
\includegraphics[width=0.8\textwidth]{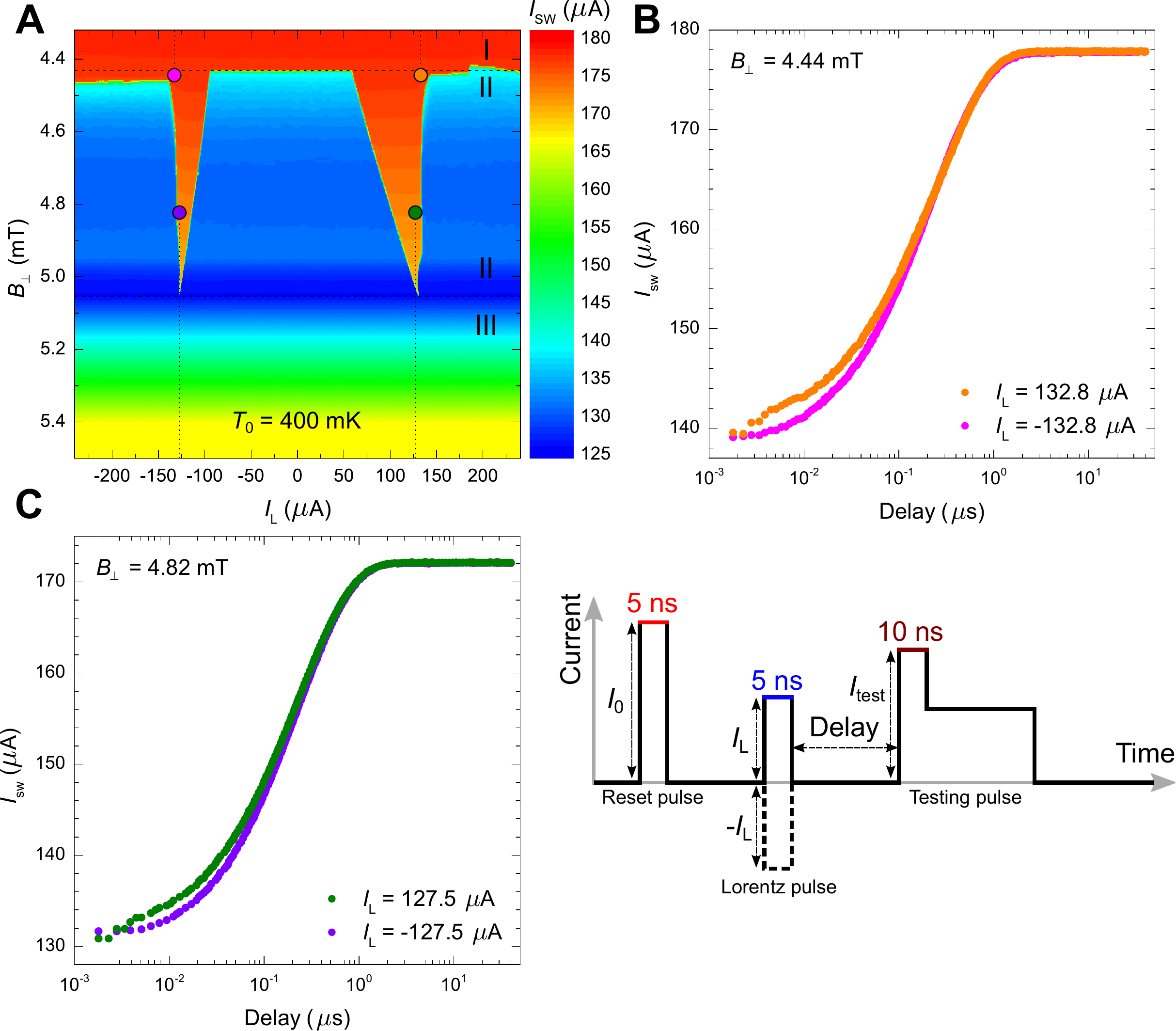}
\label{Fig_EG}
\end{figure}

\noindent Figure S9: \textbf{Thermal dynamics of the SVB after vortex expulsion for the two polarities of the Lorentz pulse.} (\textbf{A}) Vortex stability diagram, measured at 400$\,$mK. We indicate the two pairs of points, for which we record the thermal relaxations after expulsion of the vortex. (\textbf{B}) Relaxations of the switching current, measured at the two symmetric points of the vortex stability diagram ($I_L=132.8\,\mu$A and $I_L=-132.8\,\mu$A at $B_{\bot}=4.44\,$mT). Although the expulsion currents for the two polarities are significantly different (60$\,\mu$A vs. 120$\,\mu$A for positive vs. negative polarity respectively) the measured dissipation is the same, as indicated by the same suppression of the switching current. The vortex expulsion brings about $\Delta T\sim250\,$mK temperature rise of the box in both cases. (\textbf{C}) Relaxations of the switching current measured at $I_L=127.5\,\mu$A and $I_L=-127.5\,\mu$A at $B_{\bot}=4.82\,$mT. Here, the expulsion currents for the two polarities are comparable, but significantly larger than for data presented in panel B. Nevertheless, the magnitude of dissipation revealed by all relaxation profiles remains very similar.

\bibliographystyle{Science}

\section*{Acknowledgments}
The authors thank Leo Kouwenhoven, Dmitry Golubev, Milorad Milosevic and Grzegorz P. Mazur for helpful discussions.
\\
\\
\textbf{Funding:} This work is financed by the Foundation for Polish Science project ``Stochastic thermometry with Josephson junction down to nanosecond resolution'' (First TEAM/2016-1/10) and National Science Centre Poland project ``Thermodynamics of nanostructures at low temperatures'' (Sonata Bis-9, No. 2019/34/E/ST3/00432). We acknowledge the support of the European Microkelvin Platform (EMP, No. 824109 EU Horizon 2020) and EU COST Action CA21144 SUPERQUMAP.
\\
\\
\textbf{Author contributions:} M.Z. conceived the idea of the experiment and acquired the funding for the research. M.F. prepared the sample. M.F.,K.N. and M.Z. collected the data. M.F.,K.N.,A.S. and M.Z. analyzed the data. A.S. advised on the technical aspects of the experiment. M.Z. led the research and wrote the manuscript, with the input from M.F.
\\
\\
\textbf{Competing interests:} The authors declare no competing interests.

\end{document}